\documentclass[reprint,footinbib,amsmath,amssymb,aip,jcp,%
  floatfix,superscriptaddress,a4paper,nolongbibliography]{revtex4-2}
\usepackage{newtxtext,newtxmath}
\usepackage{microtype}
\usepackage{gensymb}
\usepackage[colorlinks=true,allcolors = blue]{hyperref}
\usepackage{color}
\usepackage{graphicx}

\newcommand{\dd}{\mathrm{d}}

\newcommand{\Reff}{R_{\text{eff}}}

\newcommand{\Eq}[1]{Eq.~\eqref{#1}}
\newcommand{\Eqs}[1]{Eqs.~\eqref{#1}}
\newcommand{\Fig}[1]{Fig.~\ref{#1}}

\newcommand{\Refcite}[1]{Ref.~\onlinecite{#1}}
\newcommand{\Refscite}[1]{Refs.~\onlinecite{#1}}

\newcommand{\Sec}[1]{Sec.~\ref{#1}}
\newcommand{\Appendix}[1]{Appendix~\ref{#1}}

\newcommand{\partFig}[2]{Fig.~\hyperref[#1]{\ref*{#1}#2}}

\newcommand{\german}[1]{{\itshape #1}}
\newcommand{\ansatz}{\german{ansatz}}
\newcommand{\ansatze}{\german{ans\"atze}}

\newcommand{\latin}[1]{{\itshape #1}}

\newcommand{\via}{\latin{via}}

\newcommand{\etal}{\latin{et al.}}

\newcommand{\eg}{e.g.}
\newcommand{\ie}{i.e.}
\newcommand{\cf}{cf.}

\newcommand{\code}[1]{{\sc #1}}
\newcommand{\DLMESO}{\code{dl\_meso}}
\newcommand{\UMMAP}{\code{ummap}}

\newcommand{\kB}{k_{\mathrm{B}}}
\newcommand{\kT}{\kB T}
\newcommand{\ex}{\mathrm{ex}}
\newcommand{\id}{\mathrm{id}}
\renewcommand{\textcite}[1]{\citeauthor{#1} \cite{#1}}
\newcommand{\Id}{\mathbb{I}}

\newcommand{\transpose}{\mathsf{T}}
\newcommand{\order}[1]{\mathcal{O}{\left(#1\right)}}

\newcommand{\Hq}{{\hat H}}
\newcommand{\Cq}{{\hat C}}
\newcommand{\Wq}{{\widehat W}}

\newcommand{\Omegaq}{{\hat \Omega}}

\newcommand{\myvec}[1]{{\mathbf{#1}}}
\newcommand{\rvec}{\myvec{r}}
\newcommand{\kvec}{\myvec{k}}
\newcommand{\qvec}{\myvec{q}}

\newcommand{\drvec}{\dd^3\rvec} 
\newcommand{\dqvec}{\dd^3\qvec} 

\newcommand{\convolve}{*}

\begin{document}

\title{Approximate additivity in the solvent-mediated potential of mean force for ultrasoft particle systems}

\author{Joshua F. Robinson}
\email{joshua.robinson@stfc.ac.uk}
\affiliation{The Hartree Centre, STFC Daresbury Laboratory, Warrington, WA4 4AD, United Kingdom}
\affiliation{H.\ H.\ Wills Physics Laboratory, University of Bristol, Bristol BS8 1TL, United Kingdom}

\author{Gary Yu}
\affiliation{The Hartree Centre, STFC Daresbury Laboratory, Warrington, WA4 4AD, United Kingdom}

\author{Patrick B. Warren}
\email{patrick.warren@stfc.ac.uk}
\affiliation{The Hartree Centre, STFC Daresbury Laboratory, Warrington, WA4 4AD, United Kingdom}
\affiliation{SUPA and School of Physics and Astronomy, The University of Edinburgh, Peter Guthrie Tait Road, Edinburgh EH9 3FD, United Kingdom}

\date{\today}

\begin{abstract}
  Recently an approximate additivity principle has been successfully applied to molecular chemical potentials in ultrasoft particle systems [R.\ L.\ Hendrikse, C.\ Amador and M.\ R.\ Wilson, Phys.\ Chem.\ Chem.\ Phys.\ {\bf27}, 1554 (2025)].
  We articulate this additivity principle as one of two \ansatze\ and apply these to dimers in the infinite dilution limit.
  In this limit, we show that the solvent-mediated potential of mean force (PMF) between solutes, extracted from the hypernetted-chain (HNC) closure of the Ornstein-Zernike equations, can be expressed as a convolution between solute-specific generalised excluded volume functions.
  In the limit of a structureless solvent of point particles and hard core solutes, this recovers the exact Asakura-Oosawa depletion potential as the overlap between excluded volume spheres.
  The theory extends to the opposite limit of ultrasoft particle solvents and solutes, such as those encountered in dissipative particle dynamics (DPD), where the solvent-mediated PMF can be recovered with considerable accuracy.
  These results confirm that in coarse-grained \emph{molecular} DPD simulations the parametrisation of the solute-solvent interactions is sensitive to the intramolecular bond lengths if they are smaller than the range of the solvent-solvent interaction potential, due to the overlap of the soft excluded volume functions.
\end{abstract}

\maketitle

\section{Introduction}
Extensivity is an asymptotic property of thermodynamic quantities, emerging from the effective independence of subsystems over sufficiently large length-scales.  A precise formulation of extensivity can be made in terms of \emph{additivity} of subsystems.  A familiar example is the total volume formed by the union of several bodies $B_i \subseteq \mathbb{R}^3$, which obeys an inclusion / exclusion principle:
\begin{equation}\label{eq:inclusion-exclusion}
  \sum_i V(B_i)\, - \sum_{i < j} V(B_i \cap B_j)
   \,+\!\! \sum_{i < j < k} V(B_i \cap B_j \cap B_k) \,- \,\cdots\,.
\end{equation}
In combination with continuity and rigid-motion invariance, this provides a rigorous definition of extensivity.\cite{klain_1997}
In three dimensions, these properties are held by four fundamental measures: volume, surface area, and the integrated mean and Gaussian curvatures.\cite{klain_1997}
Extensive thermodynamic variables like entropy are expressible as a weighted sum of these measures in the thermodynamic limit.
Mesoscopic interactions may become extensive where there are large size-asymmetries.
For example, depletion between large colloidal particles is captured well by a volume-additive form, as in \Eq{eq:inclusion-exclusion}, in the celebrated Asakura-Oosawa (AO) model,\cite{asakura_1954, *asakura_1958} and its extensions.\cite{robinson_2019}

Extensivity is required for the thermodynamic limit to be well-defined, but there is no general requirement that thermodynamic variables are even approximately additive on microscopic length-scales due to correlations.
Striking deviations from additivity are prominent, for example, in the interactions between solutes surrounded by wet films due to the formation of liquid bridges;\cite{stark_2004, archer_2005} these non-additive contributions may be central to droplet stabilisation in the so-called `ouzo effect'.\cite{archer_2024,*archer_2026}
This example illustrates the complete breakdown of additivity near the gas-fluid phase boundary, but far from this transition additivity may be permitted as an asymptotic or approximate property.
Additivity underlies some of the most successful (approximate) theories for hard particle systems.\cite{reiss_1959, rosenfeld_1988, *rosenfeld_1989, roth_2010}
In hard spheres additivity captures most contributions to the free energy by transforming the generally many-body free energy functional into a simple convolution of (additive) one-body functions.
This seems to occur due to the pair potential being geometric in nature.
The standard treatment of simple liquids incorporates attractions as weak perturbation around hard cores,\cite{weeks_1971, hansen_2006} so additivity can be thought to dominate the fluctuations of simple liquids.

We are motivated here by the recent work of Hendrikse~\etal\cite{hendrikse_2025} who developed a technique to systematically coarse-grain molecular liquids by assuming the chemical potential of a coarse-grained molecule is additive in its constituent parts (`beads').
They develop their parameterisations for dissipative particle dynamics (DPD) where the interaction potential is `ultrasoft' (meaning bounded, even at complete overlap).\cite{groot_1997}  In DPD the fluid is replaced by beads representing 1-6 atoms which interact through soft potentials, together with a pairwise momentum-conserving thermostat to preserve hydrodynamics.

To the best of our knowledge, the only theories that have been built on strict additivity outside of hard-core systems are Refs~\onlinecite{schmidt_1999, groh_2001} for the free energy functional, but no extension has been made to chemical potentials.
Following the lead of Saathoff,\cite{saathoff_2018} Hendrikse~\etal\cite{hendrikse_2025} apply an inclusion / exclusion relation along the lines of \Eq{eq:inclusion-exclusion} to molecular chemical potentials in terms of the individual bead chemical potentials by assigning a hard core radius to each bead, analogously to the AO model.\cite{asakura_1954, *asakura_1958}  Their numerical results were favourable to this theory despite the apparent contradiction between this `hard core' assumption and the inherent softness of DPD beads.
This is suggestive that an approximate additivity applies to the solvation free energy in soft potentials.
If correct, this property could provide a route to treat complex molecular liquids \emph{theoretically} thereby circumventing expensive numerical routes to coarse-graining (see \eg\ \Refcite{anderson_2017}).

Our aim is to investigate the additivity properties of solvent-mediated interactions in liquids characterized by ultrasoft non-bonding pair potentials, typified by DPD.
We thus focus on the $n$-particle correlation functions,
\begin{equation}
g_{\kvec} = \exp{( -\beta \phi_{\kvec} - \beta W_{\kvec})}\,,
\end{equation}
where $\kvec = (k_1, \cdots, k_n)$ indexes the atoms in the $n$-mer, $\phi_{\kvec}$ is the intramolecular interaction potential and $W_{\kvec}$ is the solvent-mediated potential.
If taken literally, Hendrikse \etal\ would essentially suggest that a potential of AO-type applies for solvent-mediated part of this.
For dimers this would imply
\begin{equation}\label{eq:w-morphometric}
  \frac{W_{ij}(r)}{W_{ij}(0)}
  =\Bigl\{\begin{array}{cl}
      \>(1+x/4)(1-x/2)^2 & (x<1)\,,\\
      \>0 & (x\ge1)\,,
  \end{array}
\end{equation}
where $x=r/R_{ij}$ and $R_{ij} = R_i + R_j$ is the combined volume-excluding radius for interactions between species $i$ and $j$ which are assumed to interact spherically.  This relationship emerges from volume additivity, as in \Eq{eq:inclusion-exclusion}, truncated pairwise.

One issue with applying the AO form in \Eq{eq:w-morphometric} to soft potentials is that it predicts a discontinuity in the force at the origin: $W_{ij}'(0)/W_{ij}(0) = -3/(4R_{ij})$.  In general, the solvent-mediated interaction should inherit the smoothness and regularity of the underlying pair potential so $W_{ij}'(0) \ne 0$ is only expected / allowed for hard core interactions where it is singular. The excellent numerical prediction of partition coefficients found by Hendrikse~\etal\ is nevertheless convincing evidence that the AO form in \Eq{eq:w-morphometric} holds approximately in soft systems.  Our aim is to develop a self-consistent liquid-state theory for $W_{\kvec}$ which contains the additivity property but without pathologies at full overlap.

The structure of the paper is as follows.
In \Sec{sec:additivity} we postulate a generalised additive form for the solvent-mediated interactions, and deduce some consequences for the confluent limits for these for ultrasoft potentials.
We show that the hypernetted chain (HNC) approximation in liquid state theory in the limit of an infinite dilution furnishes an additive 2-body form for solvent-mediated potential of mean force (PMF) which reduces to AO in the limit where interactions become hard and the solvent particles become points.
We develop an approximate form of the solvent-mediated PMF in the spirit of geometric approach of Hendrikse \etal, generalizing their volume-exclusion form in \Eq{eq:w-morphometric}.
We show numerical results in \Sec{sec:w-results}.
Finally we apply this formalism to calculate the partition coefficients of DPD dimers in monomeric solvents.

\section{Additivity in the solvent-mediated potential of mean force}\label{sec:additivity}
\subsection{General formalism}
\label{sec:additivity-ansatz}
From potential distribution theorem,\cite{widom_1963,widom_1982,robinson_2019} we can identify $W_{\kvec}$ with the reversible work to bring the monomers together from infinite separation.  A formally exact decomposition is
\begin{equation}\label{eq:w-exact}
W_{\kvec}(\rvec_1,\cdots,\rvec_n) = \Delta\Omega - \sum_i \mu_i^\ex\,.
\end{equation}
where $\Delta \Omega$ is the change in the grand potential from inserting an $n$-mer in a frozen configuration, and $\{\mu_i^\ex\}$ are the excess chemical potentials of the individual beads; subtracting these ensures $W_\kvec \to 0$ as all the component beads in the $n$-mer become macroscopically separated.

As an \ansatz, let us now suppose that the solvent-mediated potential obeys a generalised `additivity' or inclusion / exclusion principle:
\begin{equation}\label{eq:w-ansatz}
  W_{\kvec} \propto - \sum_{i < j} \psi_i \convolve \psi_j\, +\!\! \sum_{i < j < k} \psi_i \convolve \psi_j \convolve \psi_k \,-\, \cdots\,,
\end{equation}
where the $\{\psi_i\}$ are generalised excluded volume functions for each bead, and we define the symmetric 3d convolutions \via\
\begin{equation*}
\begin{split}
  &(a \convolve b)(\rvec_1, \rvec_2) = {\textstyle\int}\, \drvec\, a(\rvec - \rvec_1)\, b(\rvec - \rvec_2)\,,\\[3pt]
  &(a \convolve b \convolve c)(\rvec_1, \rvec_2, \rvec_3) = {\textstyle\int}\,  \drvec\, a(\rvec - \rvec_1)\, b(\rvec - \rvec_2)\, c(\rvec - \rvec_3)\,,\\
  \end{split}
\end{equation*}
and so on for higher multiple convolutions.

One-body terms in \Eq{eq:w-ansatz} are absent because of the requirement that $W_\kvec\to0$ as the particles become macroscopically separated; this is satisfied by the convolutions provided the $\{\psi_i\}$ decay sufficiently fast.
We can also consider a similar formulation that includes one-body terms by making $\Delta \Omega$ itself additive:
\begin{equation}\label{eq:dOmega-ansatz}
  \Delta\Omega \propto \sum_i {\textstyle\int} \drvec \, \psi_i - \sum_{i < j} \psi_i \convolve \psi_j\, +\!\! \sum_{i < j < k} \psi_i \convolve \psi_j \convolve \psi_k \,-\, \cdots\,.
\end{equation}
As $\int\! \drvec \, \psi_i(\rvec) \propto \mu_i^\ex$ is required for consistency with the definition of $W_\kvec$ in \Eq{eq:w-exact}, this alternative places a tighter restriction on the definition of $\psi_i$.
It is closer in spirit to the approach taken by Hendrikse \etal, who used the monomer results to construct molecular chemical potentials.

The terms appearing on the right-hand side of each \ansatz\ in \Eqs{eq:w-ansatz} and~\eqref{eq:dOmega-ansatz} are generalised volumes.
Noting that $\Delta \Omega \simeq p \Delta V$ for large solutes where surface terms are negligible,\cite{robinson_2019} for thermodynamic consistency we thus require that the proportionality coefficient is the solvent pressure.
\Eqs{eq:w-ansatz} or~\eqref{eq:dOmega-ansatz} recover true volume-additivity in \Eq{eq:inclusion-exclusion} in the limit where the $\{\psi_i\}$ become indicator functions.  In this limit, at the pairwise level, $W_{ij}$ reduces to the AO form in \Eq{eq:w-morphometric} with the constant of proportionality being the solvent pressure (or density, for a solvent of point particles as in the AO model).
We will comment further on these morphological limits in a later section.

\subsection{Zero-separation theorems}
To assess how well the additivity \ansatz\ in \Eq{eq:w-ansatz} captures correlations in ultrasoft systems, we introduce a set of self-consistency relationships derived from zero-separation theorems.\cite{lee_1996}
These theorems express the idea that co-localizing particles is equivalent to inserting a single particle with combined pair potentials $\phi_{1+2} \equiv \phi_{1} + \phi_{2}$.  For example, assuming \Eq{eq:w-ansatz}, the volume-exclusion form for the solvent-mediated potential for trimers would entail
\begin{equation*}
\begin{split}
  &W_{123}(\rvec_1,\rvec_2,\rvec_3)
  \propto {}- \psi_1 \convolve \psi_2 - \psi_1 \convolve \psi_3 - \psi_2 \convolve \psi_3\\[3pt]
  &\hspace{6em}{}+ {\textstyle\int}\, \drvec \, \psi_1(\rvec_1 - \rvec)\, \psi_2(\rvec_2 - \rvec)\, \psi_3(\rvec_3 - \rvec)\,.
\end{split}
\end{equation*}
In the confluent limit $\rvec_2 \to \rvec_1$ this becomes
\begin{equation}
  W_{123}(\rvec_1,\rvec_1,\rvec_3) 
  \propto{}-(\psi_1 + \psi_2 - \psi_1\,\psi_2)\convolve\psi_3\,,\label{eq:w123a}
\end{equation}
where we omit the constant term coming from $\psi_1\convolve\psi_2$.
As this situation is equivalent to the dimer $(1+2, 3)$, we equate the left hand side with
\begin{equation}
  W_{1+2,3}(\rvec_1,\rvec_3) \propto -\psi_{1+2} \convolve \psi_3\,.\label{eq:w123b}
\end{equation}
But $\psi_3$ is an arbitrary test-function so we can neglect the null-space in the convolution, and read off from \Eqs{eq:w123a} and~\eqref{eq:w123b}
\begin{equation}\label{eq:additivity-self-consistency}
  \psi_{1+2} = \psi_1 + \psi_2 - \psi_1\, \psi_2\,.
\end{equation}
The same expression is obtained in identical fashion from $W_{123}$ assuming instead \Eq{eq:dOmega-ansatz}.

As another self-consistency check, we consider the co-localisation of two particles inside a dimer.
For the \ansatz\ involving one-body terms in \Eq{eq:dOmega-ansatz}, we thus find
\begin{equation*}
\Delta \Omega
=
{\textstyle\int} \drvec \left(
  \psi_1 + \psi_2 - \psi_1 \psi_2
\right)\,.
\end{equation*}
This limit entails setting $\Delta \Omega = \mu_{i+j}^\ex$ \ie\ the excess chemical potential of a monomer interacting through $\{\phi_{ik} + \phi_{jk}\}_k$.
\Eq{eq:dOmega-ansatz} also encompasses $\mu_{i+j}^\ex = \int\! \drvec \, \psi_{i+j}$, so this route leads to
\begin{equation}\label{eq:additivity-self-consistency-2}
  \psi_{1+2} = \psi_1 + \psi_2 - \psi_1\, \psi_2 + \Psi\,,
\end{equation}
where $\Psi$ is any function integrating to zero over the system volume $\int\!\drvec\,\Psi = 0$.
The latter terms have no interpretation within the geometric framework where $\{\psi_i\}$ are generalised indicator functions, and so can be dropped.
The self-consistency relation in \Eq{eq:additivity-self-consistency-2} thus becomes identical to the previous relation in \Eq{eq:additivity-self-consistency}.

Finally, we note that in the limit where all the particles are co-localised, inserting an $n$-mer becomes equivalent to inserting a monomer with a summed solute-solvent interaction potential.  Applying this to a dimer for example yields\cite{saathoff_2018}
\begin{equation}\label{eq:cavity-self-consistency}
  W_{ij}(0) \equiv \lim_{r \to 0} W_{ij}(r) = \mu_{i+j}^\ex - \mu_i^\ex - \mu_j^\ex\,.
\end{equation}
This result expresses a well-known zero-separation theorem for a pair of particles (or cavities) in a fluid.\cite{lee_1996}
It is an \emph{exact} result that can be used to assess thermodynamic self-consistency in general.
Our main result though is the relationship \Eq{eq:additivity-self-consistency}, which provides a test of the self-consistency of the specific additivity assumptions of \Eqs{eq:w-ansatz} or~\eqref{eq:dOmega-ansatz}.
In the next section we will argue for a particular form for the functions $\psi_i$ which approximately applies for ultrasoft potentials.

\subsection{Liquid state theories for solutes at infinite dilution}
Starting from the pair distribution functions $g_{ij}(r)$, it is conventional in liquid state theory to define the total correlation functions $h_{ij}(r)=g_{ij}(r)-1$, and the direct correlation functions $c_{ij}(r)$ \via\ the Ornstein-Zernike (OZ) relations,
\begin{equation}\label{eq:oz}
h_{ij} - c_{ij} = {\textstyle\sum_k} \,\rho_k \, c_{ik} \convolve h_{jk}\,,
\end{equation}
where $\{\rho_i\}$ are the bulk concentrations of each species (the combinations $h_{ij}-c_{ij}$ are then known as the indirect correlation functions).  One further defines the so-called bridge functions $b_{ij}(r)$ \via\ 
\begin{equation}\label{eq:gij-exact}
\ln g_{ij}={}-\beta \phi_{ij}+h_{ij}-c_{ij}+b_{ij}\,,
\end{equation}
where $\phi_{ij}(r)$ is the bare pair potential and $\beta=1/\kT$ is the inverse temperature.  This definition implies that the solvent-mediated PMF for dimers is given exactly by 
\begin{equation}\label{eq:wij-exact}
  \beta W_{ij} = {}- h_{ij}+c_{ij}-b_{ij}\,.
\end{equation}

We now focus on the limit of infinite dilution of all components except one, which we identify with the solvent.  
In this limit the OZ relation in \Eq{eq:oz} decouples into an autonomous equation for the solvent, and equations for the solutes coupled to the solvent.
For a monatomic solvent we label the solvent species with $0$ and set $\rho_i = 0$ for $i > 0$, yielding
\begin{equation}\label{eq:OZ-dilute}
  h_{ij} - c_{ij} 
  = \rho_0 c_{0i} \convolve h_{0j}
\end{equation}
(note that $h_{ij}=h_{ji}$, etc).
In reciprocal space, denoted by hats on the functions, the solvent-solute equation ($j = 0$) leads to
\begin{equation*}
\hat{c}_{0i}(q) 
= \frac{\hat{h}_{0i}(q)}{1+\rho_0\hat{h}_{00}(q)}\,.
\end{equation*}
Injecting this back into the solute-solute OZ equation gives
\begin{equation*}
\hat{h}_{ij}-\hat{c}_{ij} =
\rho_0 \, \frac{\hat{h}_{0i} \hat{h}_{0j}}{S_0(q)}\,.
\end{equation*}
where $S_0(q) \equiv 1 + \rho_0 \hat{h}_{00}(q)$ is the solvent structure factor.

Let us now define the single-solute functions 
\begin{equation}\label{eq:psi-def}
  \hat\psi_i(q) = - \sqrt{\frac{\rho_0}{\beta p_0 \, S_0(q)}}\, 
  \hat{h}_{0i}(q)\,.
\end{equation}
where the judicious choice of sign and prefactor will be explained below (in the limit of infinite dilution there is no difference between the system pressure $p$ and the solvent pressure $p_0$, but we retain the subscript to remind the reader that all the factors in the prefactor are associated to the pure solvent).
With these functions in hand we can write, in real space, 
\begin{equation}\label{eq:additive-indirect}
h_{ij}-c_{ij} =\beta p_0\, \psi_i\convolve \psi_j\,.
\end{equation}
We have thus found that the indirect correlation function is additive in the sense of our \ansatz\ in \Eq{eq:w-ansatz}.
Finally, we see that if the bridge function is approximately zero $b_{ij} \simeq 0$ then the solvent-mediated potential for dimers in \Eq{eq:wij-exact} becomes
\begin{equation}\label{eq:wij-general}
  W_{ij} = - p_0 \, \psi_i \convolve \psi_j\,,
\end{equation}
This expression represents our principal main result.  It expresses the solvent-mediated PMF as the convolution of two single-solute properties, and can be interpreted as the geometric analogue of the volume additivity discussed in the introduction.  It holds under the assumption that the bridge function between the two solutes can be neglected.  We have concentrated on a single component monatomic solvent for notational simplicity, but one can show that the additive form in \Eq{eq:wij-general} holds for multicomponent (but still monatomic) solvents with a suitable generalisation of the $\{\psi_i\}$ functions; see \Appendix{appendix:multi} for details.

The additive form in \Eq{eq:psi-def} is evidently consistent with the general additive \ansatz\ in \Eq{eq:w-ansatz} at the two-body level.
As we mentioned briefly in the previous section upon introducing the additivity \ansatz\ in \Eq{eq:w-ansatz}, this form generates the excluded volume in the colloid-polymer limit where the solvent is composed of point particles and the solutes are hard spheres.
This limit corresponds to a structureless solvent with $\beta p_0=\rho_0$ and $S_0(q)=1$ identically, and therefore $\psi_i=-h_{0i}$ according to \Eq{eq:psi-def}.
For hard solute-solvent interactions, $g_{0i}=\Theta(r-R_i)$ where $R_i$ is the radius of the solute and $\Theta(x)$ is the Heaviside function, and therefore $\psi_i=\Theta(R_i-r)$ vanishes outside of the hard core and $\psi_i=1$ within the hard core.
The convolution in \Eq{eq:wij-general} then selects the overlap region, making the connection with geometric approaches to depletion. 
For this reason we refer to the $\{\psi_i\}$ defined in \Eq{eq:psi-def} as generalised excluded volume functions.  This motivates the choice of sign in \Eq{eq:psi-def} and the inclusion of $p_0$ in the prefactor, so that \Eq{eq:wij-general} obtains.

Equation~\eqref{eq:wij-general} holds for solutions satisfying the OZ relations where the bridge function is zero.  In liquid state theory, this is precisely the hypernetted chain (HNC) approximation.  It is known that the HNC approximation is nearly exact for ultrasoft potentials at moderately large densities,\cite{louis_2000} as is the case with DPD.  We thus use HNC as the ground-truth for this system.  With this approach we can extract the generalised excluded volume functions $\psi_i$ in DPD by solving the OZ equation with HNC closure and applying \Eq{eq:psi-def}.

In HNC, an alternative function can also be extracted to correspond to the additivity assumption for $\Delta \Omega$ in \Eq{eq:dOmega-ansatz}.  This is because in HNC, the monomer chemical potential can be written as the one-body integral\cite{hansen_2006}
\begin{equation}\label{eq:hncmui}
  \beta \mu_i^\ex \simeq \textstyle{\sum_j} \,\rho_j
  \,{\textstyle\int} \drvec \, 
  [{\textstyle\frac{1}{2}}h_{ij} ( h_{ij} - c_{ij} )- c_{ij}]\,,
\end{equation}
without coupling to intermediate densities.
This result can be understood as central to the HNC approximation within a test-particle framework (\cf\ Appendix \ref{appendix:test-particle}).
From this we read off an alternative generalised `indicator' function, in the limit of infinite dilution and for a monatomic solvent,
\begin{equation}\label{eq:psi-one-body}
\psi_i(r) = \frac{\rho_0}{\beta p_0} 
[{\textstyle\frac{1}{2}} h_{0i} ( h_{0i} - c_{0i} ) - c_{0i}]\,.
\end{equation}
This function will in general be different to that obtained from the OZ relation in \Eq{eq:psi-def}.  As in that case 
we incorporate the pressure $p_0$ into the definition here for thermodynamic consistency (see discussion at end of \Sec{sec:additivity-ansatz}).

The outstanding question is whether the $\{\psi_i\}$ obtained via the two- and one-body routes respectively in \Eqs{eq:psi-def} and~\eqref{eq:psi-one-body} respect a more general additive principle beyond dimers.  We test this be examining whether they meet the self-consistency relationship in \Eq{eq:additivity-self-consistency} derived in the previous section.

\subsection{Heuristic approach}
In the introduction we rejected the explicit AO form in \Eq{eq:w-morphometric} suggested by Hendrikse~\etal\cite{hendrikse_2025} due to pathologies at the origin approaching complete overlap.
However, a simple heuristic form along the same lines would be advantageous for implementations.
In the spirit of finding a simple form for the potential, we propose the following empirical function:
\begin{subequations}\label{eq:fit-function}
\begin{equation}
  \frac{W_{ij}(r)}{W_{ij}(0)}
  =\Bigl\{\begin{array}{cl}
  (1+2x)(1-x)^2 & (x<1)\,,\\
  0 & (x\ge1)
  \end{array}
\end{equation}
(where $x=r/R_{ij}$ as before). 
The functional form for $x<1$ is the minimal non-trivial polynomial $f(x)$ satisfying
\begin{equation}
f'(0)=f(1)=f'(1)=0\,.
\end{equation}
\end{subequations}
Unlike the AO form in \Eq{eq:w-morphometric}, this potential has the correct property that $W_{ij}'(0) = 0$ restoring regularity.
The value of $W_{ij}(0)$ can be fixed by requiring thermodynamic self-consistency as in \Eq{eq:cavity-self-consistency} with knowledge of the monomer chemical potentials.
We will explore whether this simple form respects the second self-consistency requirement \Eq{eq:additivity-self-consistency}. 

The empirical form \Eq{eq:fit-function} is not consistent with the additive \ansatz\ in \Eq{eq:w-ansatz}.
This can be seen from the Fourier transform $\Wq_{ij}(q)$ which is sign indefinite.
There is therefore no $\hat{\psi}_i(q)$ which solves $\Wq_{ii}=-p_0[\hat{\psi}_i(q)]^2$ from inverting the additive solvent-mediated potential for dimers in \Eq{eq:wij-general}.  However, inspecting $\Wq_{ij}(q)$ suggests it may be approximated by a similar functional form in reciprocal space,
\begin{equation}\label{eq:wijq-approx}
  \frac{15}{4\pi R_{ij}^3}\frac{\Wq_{ij}(q)}{W_{ij}(0)}
  =\Bigl\{\begin{array}{cl}
  (1+2y)(1-y)^2 & (y<1)\,,\\
  0 & (y\ge1)\,,
  \end{array}
\end{equation}
with the normalisation being such that $\Wq_{ij}(q\to0)$ comes out correctly, and $y=\alpha qR_{ij}/2\pi$ where $\alpha=\order{1}$ is a suitably chosen scale factor.  
To fix the latter we impose that $(2\pi)^{-3}\!\int \dqvec\, \Wq_{ij}(q)=W_{ij}(0)$ in \Eq{eq:wijq-approx} which yields $\alpha=(4\pi/15)^{2/3}\simeq 0.889$.  The functional form in \Eq{eq:wijq-approx} is positive-definite and so the square root can be taken to yield a heuristic for $\hat\psi_i(q)$, which could be taken forward to estimate the convolutions in \Eq{eq:w-ansatz} for instance.  The advantage of this approach over HNC is that the exact limiting value $W_{ij}(0)$ from the zero-separation theorem in \Eq{eq:cavity-self-consistency} can be used.

\section{Results for the solvent-mediated potential}\label{sec:w-results}
\subsection{DPD potential}
For non-bonded interactions we will focus on the standard short-range repulsive potential used in DPD:
\begin{equation}\label{eq:phi-dpd}
\frac{\beta \phi_{ij}(r)}{A_{ij}} =\Bigl\{\begin{array}{cl}
(1-x)^2/\,2 & (x<1)\,,\\
0 & (x\ge1)\,,
\end{array}
\end{equation}
where as before $x=r/R_{ij}$, and $A_{ij}$ is a species-dependent repulsion amplitude.
It is convenient to simplify this by using the cutoff as the unit of length $R_{ij} = 1$.
The water-like solvent in DPD typically has interaction strength $A_{00} = 25$ with density $\rho_0 = 3$.\cite{groot_1997}
In DPD the bead radius is normally scaled by mass, and so the radii tends to be varied less than the repulsion amplitudes.
For this reason we will work with solute beads which also have $R_{0i} = 1$ but vary $A_{0i}$.

\subsection{Comparison with simulation}

\begin{figure}
  \centering
  \includegraphics[width=\linewidth]{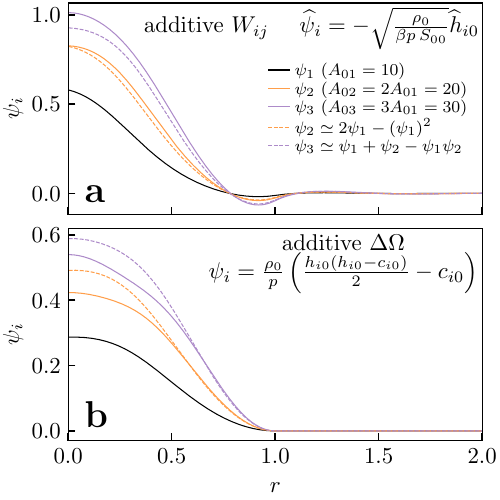}
  \caption{
    Generalised indicator functions extracted by the two methods described in the text: (a) the `two-body' route of \Eq{eq:psi-def} assuming the \ansatz\ of \Eq{eq:w-ansatz} and (b) the `one-body' route of \Eq{eq:psi-one-body} assuming the \ansatz\ of \Eq{eq:dOmega-ansatz}.
    The solid lines show the indicator functions extracted from the route, whereas the dashed lines show the expected result from the additive self-consistency relationship \Eq{eq:additivity-self-consistency}.
  }
  \label{fig:indicators}
\end{figure}

In \Fig{fig:indicators} we show the functions $\psi_i$ extracted by the methods outlined in the previous section.
\Fig{fig:indicators}(a) shows $\psi_i$ obtained from the `two-body' route with \Eq{eq:psi-def} whereas \Fig{fig:indicators}(b) shows that obtained via the `one-body' route with \Eq{eq:psi-one-body}.
Both sets of functions (solid lines) resemble generalised indicator functions as expected: they are short-ranged and decaying.
However, the functions obtained by the two-body route stretch this geometric interpretation somewhat by not being strictly monotone and featuring a small region of negative $\psi_i$ approaching $r \to 1$ from below.
The one-body functions though are strictly monotone with $\psi_i > 0$.

We see good self-consistency with respect to the additivity \ansatz.
\Fig{fig:indicators} shows the expected functions (dashed lines) obtained by application of the zero-separation-derived result \Eq{eq:additivity-self-consistency}.
These approximate functions (dashed lines) compare favourably with the `exact' results extracted by each method (solid lines).
Small deviations are seen at small $r$, but we see excellent agreement in the tails for $r \gtrsim 0.4$.
This agreement is suggestive that correlations in this system are approximately additive.
Hendrikse \etal\ considered molecules with separations $r \in \{0.4, 0.6, 1\}$ inside the region where additivity is quantitatively accurate.

\begin{figure}
  \centering
  \includegraphics[width=\linewidth]{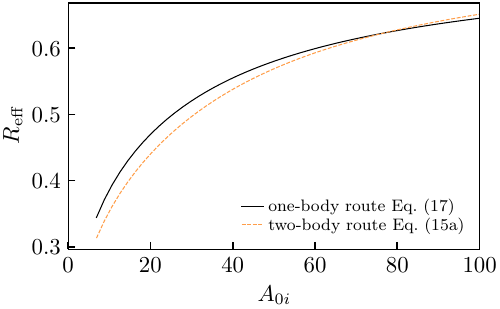}
  \caption{Effective hard-sphere radius calculated from \Eq{eq:Reff}.}
  \label{fig:effective-radius}
\end{figure}

We can define an equivalent hard sphere radius as
\begin{equation}\label{eq:Reff}
\textstyle{\frac{4}{3}}\pi\Reff^3 = {\textstyle\int}\drvec\,\psi_i(r)\>[{}=\hat\psi_i(0)\,]
\end{equation}
(this is literally the excluded volume implied by these generalised excluded volume functions).
Approaching the solvent interaction $A_{0i} \simeq A_{00}$ we find that $R_{\text{eff}} \simeq 0.5$.
In \Fig{fig:effective-radius} we show that $R_\text{eff}$ typically varies from $\simeq0.4$ to $0.6$ in the region $10\le A_{0i}\le100$, so $R_\text{eff} \simeq 0.5$ is a reasonable approximation.
This corresponds to an effective hard sphere diameter of $2 R_\text{eff} \simeq 1$ which is simply our assumed DPD interaction length.
We note that Hendrikse \etal\ used the DPD interaction length in their geometric construction, so our analysis provides a justification for their geometric construction even though we have assumed a softer notion of additivity.

\begin{figure}[t]
    \centering
    \includegraphics[width=\linewidth]{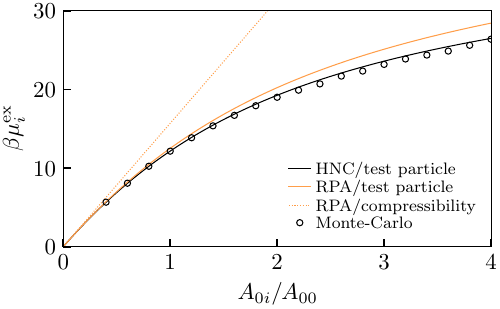}
    \caption{
        Excess chemical potential of monomers inserted into a water-like DPD solvent with $\rho_0 = 3$ and $A_{00} = 25$.
        Theoretical approaches (lines) are described in text.
        The circles are Monte-Carlo simulation data from Hendrikse~\etal,
        reproduced from \Refcite{hendrikse_2025} with permission from the PCCP Owner Societies.
    }
    \label{fig:monomer-mu}
\end{figure}

Our main results above are suggestive that the additivity principle is at play in ultrasoft DPD systems.
However, they rest on the laurels of the HNC providing the `ground truth' for this system.
In the remainder of the section we provide numerical evidence that the HNC is numerically quasi-exact.

In \Fig{fig:monomer-mu} we show that HNC is near-exact for the monomer chemical potential for water beads in DPD when compared to the Monte-Carlo data from Hendrikse~\etal\cite{hendrikse_2025}.
Ultrasoft DPD particles form a nearly mean-field system: the moderately high density $\rho_0 = 3$ ensures each particle has a large number of neigbours, but combined with the moderate interaction strength $A_{00} = 25$ induces weak deviations from the strict limit.
The mean-field result from the random-phase approximation (RPA) is linear in solvent-solute interaction strength $A_{0i}$ (dotted orange line in \Fig{fig:monomer-mu}) which only works at small $A_{0i} / A_{00}$.
HNC captures non-linear deviations from this result exceedingly well at moderate $A_{0i} / A_{00} = \order{1}$.
We note that much larger values of $A_{01}$ would involve stronger hard-sphere-like deviations from mean-field, which would not captured by the HNC.
We can thus proceed with HNC so long as $A_{01} / A_{00} = \order{1}$, which is not so restrictive.

Curiously, the less accurate RPA is not additive.
The standard RPA approach still imposes the OZ equation, and so we still obtain the additive indirect correlation of \Eq{eq:additive-indirect}.
However, this approach discards the closure for the solvent-mediated potential \Eq{eq:wij-exact} and so we instead obtain
\begin{equation*}
  W_{ij}
  =
  -\ln{\left( 1 - \beta \phi_{ij} + \beta p_0 \, \psi_i \ast \psi_j \right)} - \beta \phi_{ij}
\end{equation*}
directly from the OZ equation.
This is \emph{not} of additive form, suggesting any putative volume-additivity emerges outside of the mean-field limit.
This raises the question of how additive related theories are.
HNC can be well-understood within the test-particle framework (\cf\ Appendix \ref{appendix:test-particle} for details), and related theories can be obtained in this framework.  Archer and Evans\cite{archer_2003} studied the `test-particle RPA' (TP-RPA) which is loosely a hybrid of HNC and RPA.
This theory does not satisfy the the OZ equation or the closure \Eq{eq:wij-exact}, and so there is no suggestion of additivity either despite its structural similarity to the HNC.
This theory captures some of the non-mean-field corrections (dashed orange line in \Fig{fig:monomer-mu}) but not quite as accurately as HNC.

\begin{figure}
  \centering
  \includegraphics[width=\linewidth]{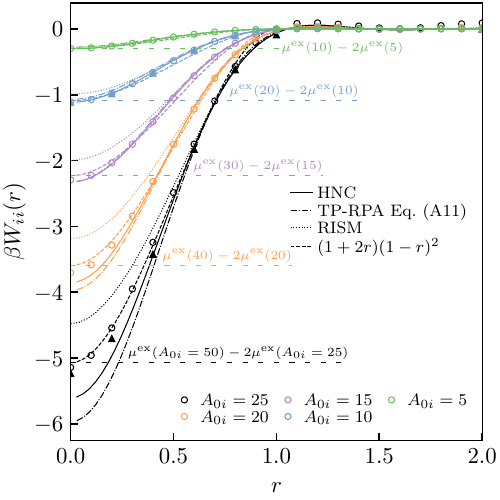}
  \caption{Solvent-mediated potential between a pair of identical solutes (`1') in a solvent of DPD water beads (`0'), as the solute-solvent interaction parameter $A_{01}$ is varied.
  Rigid dimer chemical potentials from the various theories (lines) and measured by Widom insertion into a $50\times50\times50$ box (triangles) and $10\times10\times10$ box (circles).
  The horizontal dashed lines show the expected value at $r = 0$ consistent with the zero-separation theorem \Eq{eq:cavity-self-consistency} using HNC to estimate the monomer chemical potentials.}
  \label{fig:W11}
\end{figure}

HNC also accurately captures the solvent-mediated PMF measured in simulation.
We access this by measuring $\Delta \Omega_{ij}(r)$ directly by Widom insertion of rigid dimers of length $r$,
separately measuring the monomer chemical potentials also by Widom insertion.
The solvent-mediated PMF then follows naturally from the theory via \Eq{eq:w-exact}, 
For these simulations we equilibrated the standard DPD water model  ($A_{00} = 25$, $\rho = 3$) in $10\times10\times10$ boxes using bespoke Monte-Carlo methods, and a $50\times50\times50$ box using \DLMESO.\cite{seaton_2013}  We then performed trial insertions of fixed-length homodimers with the desired solvent-solute repulsion amplitude $A_{01}$, using a modified version of \UMMAP\ to analyse the \DLMESO\ simulation trajectories.\cite{bray_2020}
\Fig{fig:W11} shows the dimer chemical potentials from these compare favourably with the solvent-mediated PMFs calculated through all theoretical routes at small $A_{01}$.
The error is largest approaching $r \to 0$ where the theories show small violations of the exact zero separation theorem of \Eq{eq:cavity-self-consistency}.
We note that small separations are also where we saw the additivity \ansatze\ perform worse, which may be an amplification of this breakdown in self-consistency.

The theories without built-in additivity (RPA and TP-RPA) are quantitatively less accurate when compared to the Monte-Carlo data.
We have not shown the result of direct calculation with RPA because it performs poorly and clutters the plot.
We follow Archer and Evans\cite{archer_2003} by writing $W_{ij}$ for TP-RPA using the one-body direct correlation function (\cf\ \Eq{eq:wij-rpa} in the Appendix).
This results (dash dotted line in \Fig{fig:W11}) in a less accurate solvent-mediated potential  is less accurate than the latter direct approach (dash dot) and the HNC approach.
The relative increases at small $r$ with this theory being less self-consistent with respect to the exact zero separation theorem \Eq{eq:cavity-self-consistency} than in HNC.
We speculate RPA and TP-RPA are less accurate due to lacking a built-in additivity principle.

The empirical theory of \Eq{eq:fit-function} offers near-perfect accuracy for $W_{ii}$ in the region $r \in [0, 1]$ (dashed line).
This theory requires the excess monomer chemical potential as input, which we took from the HNC result.
Its improvement in accuracy over the HNC approaching $r \to 0$ despite a lack of additivity contradicts our observation that non-additive theories are less accurate.
However, this comes at the price of the tails of $W_{ii}$ for $r > 1$ which are completely neglected. 

The principal advantage of the additive expression for $W_{ij}$ is that it provides a route to polyatomic molecules beyond dimers.
The standard liquid state theory approach to treat such molecules is the reference interaction site model (RISM).
For comparison we implemented the RISM approach for DPD particles pioneered by Fraaije~\etal,\cite{fraaije_2016} as described in Appendix~\ref{appendix:rism}.
The resulting $W_{ii}$ (dotted line in \Fig{fig:W11}) performs poorly by comparison with the other theories.
This theory relies on a particular assumption about the intramolecular correlations that could in principle be fine-tuned to improve the theory, but a systematic approach to refine this would be significantly more challenging than using generalised volume additivity.

\section{Application to partitioning of dimers between solvent phases}
\label{sec:partition}

\subsection{Chemical potential of non-rigid dimers}
\label{sec:non-rigid}

For non-rigid dimers we must introduce an intramolecular bonding potential.
For consistency with Hendrikse~\etal\cite{hendrikse_2025} we assume Hookean springs,
\begin{equation}\label{eq:phi-dpd-bonded}
  \beta \phi_{ij}^\mathrm{b}(r) = k ( r - l_0 )^2\,.
\end{equation}
Hendrikse~\etal\cite{hendrikse_2025} consider an additional three-body bending potential (required to model complex molecules), but as we focus on dimers we only need \Eq{eq:phi-dpd-bonded}.

Dimers can be modelled within an atomistic framework by integrating the pair density $\rho_{ij} \equiv \rho_i \rho_j g_{ij}$ over all configurations and fixing $\rho_i = \rho_j = \rho_{\text{dim}}$ to obtain the total number of dimers.
We find their concentration by dividing through by $V$: \cite{robinson_2019}
\begin{equation}\label{eq:dimer-concentration}
C_{ij} = \frac{\rho_i \rho_j}{V} {\textstyle\int} \drvec\, \drvec' \, g_{ij}(|\rvec - \rvec'|)\,.
\end{equation}
Defining $C_{ij} \equiv \rho_i \rho_j \exp{(\beta \mu_{ij}^\ex)} = \rho_{\text{dim}}^2 \exp{(\beta \mu_{ij}^\ex)}$, we write the dimer's excess chemical potential in the limit of infinite dilution as
\begin{equation}\label{eq:mu-dimer}
\begin{split}
\beta\mu_{ij}^\ex &=
\beta\mu_i^\ex + \beta\mu_j^\ex\\[3pt]
&\qquad{}+ \ln{ {\textstyle\int} \drvec \, \exp[{-\beta (\phi_{ij}^\text{b}(r) + \Delta \Omega_{ij}(r) )] } }\,.
\end{split}
\end{equation}
Rigid dimers are recovered in the $k\to\infty$ limit.

\subsection{Evaluating partition coefficients between monoatomic solvent phases}

\begin{figure}
  \centering
  \includegraphics[width=\linewidth]{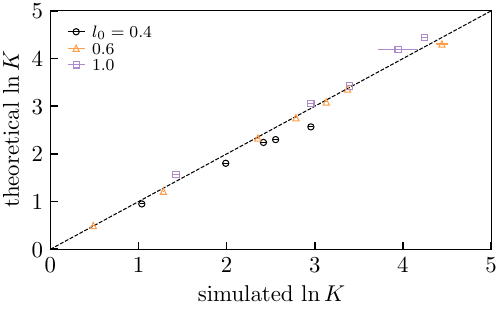}
  \caption{
    Partition coefficients of ultrasoft homodimers in a monoatomic solvents.
    Simulation data from Hendrikse~\etal\cite{hendrikse_2025} compared with the theoretical prediction of HNC via \Eq{eq:wij-general}.
    Colours represent the equilibrium dimer separation for the assumed bond potential of \Eq{eq:phi-dpd-bonded} with spring constant $k=150$.
    Within each dataset each point represents a unique pair of solute-solvent interactions for each solvent phase.
    Dashed line: ideal 1:1 agreement as a guide to the eye.  
    Simulation data reproduced from \Refcite{hendrikse_2025} with permission from the PCCP Owner Societies.
   }
  \label{fig:lnK}
\end{figure}

The partition coefficient characterises the preferential behaviour for a molecule to solvate in one of two phases.   For a given molecule (here labelled $i$), 
it is defined as the limiting ratio of the molecular concentrations in two phases $A$ and $B$ at infinite dilution:
\begin{equation}
  K_i \equiv \frac{C_i^{(A)}}{C_i^{(B)}} = \exp{[ \beta ( \mu_i^{\ex, (B)} - \mu_i^{\ex, (A)} )]}\,,
\end{equation}
where $C_i^{(\alpha)}$ are the concentrations and $\mu_i^{\ex,(\alpha)}$ are the excess chemical potentials in each phase $\alpha \in \{A, B\}$.
Typically $A$ and $B$ are chosen to be equilibrated phases of water and oil (normally 1-octanol), so that $K_i$ quantifies the molecule's relative hydro- and lipophilicity.
Parameterisations of coarse-grained systems based on this metric have proven successful outside of their training set.\cite{anderson_2017}
The downside of this approach is that direct measurement of the partition coefficients via simulation is expensive.
The present theory permits efficient calculation for monomers and dimers at least in monomeric solvents.

For monomers the partition coefficient is trivial as the chemical potentials $\mu_i^{\ex,(A)}$ and $\mu_i^{\ex,(B)}$ are straightforward to calculate by solving the OZ relation in \Eq{eq:oz} with HNC closure and using the usual expression as in \Eq{eq:mu-tp}.
The phase $\alpha$ is specified by distinct $\{A_{00}, \rho_0\}$ parameters for DPD and the resulting correlation functions \eg\ $h_{00}$.
For non-rigid dimers, their overall concentration is found by integrating the pair density over the separation using \Eq{eq:dimer-concentration}.
The partition coefficient for the dimer is then calculated as
\begin{equation}
K_{ij}
\equiv
\frac{C_{ij}^{(A)}}{C_{ij}^{(B)}}
=
K_i K_j
\frac{
\int\!\drvec\, \exp{\left( - \beta \phi_{ij}^\text{b}(r) - \beta W_{ij}^{(B)}(r) \right)}
}{
\int\!\drvec\, \exp{\left( - \beta \phi_{ij}^\text{b}(r) - \beta W_{ij}^{(A)}(r) \right)}
}
\,.
\end{equation}
For $k \to \infty$ (rigid dimers of fixed length) this reduces to
\begin{equation*}
  K_{ij}
  =
  K_i K_j \exp{[ \beta (W_{ij}^{(A)}(l_0) -  W_{ij}^{(B)}(l_0) )]}\,.
\end{equation*}
For consistency with Hendrikse~\etal\cite{hendrikse_2025} we assume a spring constant of $k = 150$ which requires integration over all realisations of the dimer.

In \Fig{fig:lnK} we show how the HNC-calculated partition coefficients compares with simulation values from Hendrikse~\etal\
The theory provides excellent agreement with the simulations.

\section{Discussion and conclusion}
Taken together, our results provide a justification for the geometric construction of Hendrikse~\etal\ so long as moderate interaction strengths are assumed ($A_{0i} = \order{10}$) and bead separations are sufficiently far from full overlap ($r \gtrsim 0.5$).
We have shown that the effective hard sphere diameter in this regime is approximately the same as the interaction length, providing a regime of validity for Hendrikse~\etal's geometric construction.
Outside of this regime our more general theory involving `soft' volumes must be deployed.

The additivity principle we have explored in this manuscript is closely connected to theories of hard particle systems.
Geometric constructions are natural for hard particle systems, and unexpected (at least by us) for ultrasoft systems so the connection warrants discussion.
Fundamental measure theory (FMT) \cite{rosenfeld_1988, *rosenfeld_1989, roth_2010} and its extensions (see \eg\ \Refscite{schmidt_1999, groh_2001, wittmannEPL2015, *wittmann_2016} and references therein) express the free energy density in terms of intrinsic volumes (related to the excluded volumes, surface areas and curvatures).
The latter quantities (called \emph{fundamental measures} in FMT) are strictly additive quantities which obey inclusion / exclusion principles much like in our \ansatze\ in \Eqs{eq:w-ansatz} and~\eqref{eq:dOmega-ansatz}.
FMT has been generalised to ultrasoft star polymers with great numerical accuracy,\cite{schmidt_1999, groh_2001} reinforcing our core message that morphological theories can be applied in such systems.
We note that our argument for an additive potential from HNC relies on neglecting the bridge function.
The bridge function is also implicitly neglected in FMT, which can be seen from its correspondence with the Percus-Yevick closure to the OZ equation.\cite{rosenfeld_1988, *rosenfeld_1989}
It is possible that a negligible bridge function is necessary for additive contributions to be seen over correlation length scales.

Closely related to FMT is the morphometric approach,\cite{konig_2004, robinson_2019} where the chemical potential of a complex hard solute is expressed as a linear function of the intrinsic volumes (\ie is additive).
This emerges as a limit of FMT, and arguably was anticipated much earlier in scaled particle theory.\cite{reiss_1959}
Our \ansatze\ correspond to the leading volume term of the morphometric approach in the limit of hard particles (the AO potential).
Borrowing intuition from FMT, we can imagine a generalised \ansatz\ to involve four generalised weight functions for the volume, surface and integrated curvatures (including vector forms for the surface functions).
Such a theory would essentially be a generalisation of the morphometric approach.
The additional degrees of freedom in this more general theory would potentially be able to unify the two similar approaches we have explored, working with the \ansatze\ in \Eqs{eq:w-ansatz} and~\eqref{eq:dOmega-ansatz}, so that both the monomer chemical potential \Eq{eq:hncmui} and the dimer OZ \Eq{eq:OZ-dilute} can be imposed with a single indicator $\chi_i$.

A more comprehensive investigation of all OZ closures\cite{pihlajamaa_2024} could in principle yield better results, because this system is near-mean-field which \emph{should} be covered by those selected here.
We suspect to go beyond this one would need many-body terms in the free energy.
A promising new approach could be to apply machine learned free energy functionals using neural networks;\cite{kampa_2025, *kampa_2026} such free energies would lead to numerically exact results that would enable more efficient parameterisations of coarse-grained systems.

\begin{acknowledgments}
JFR and PBW would like to thank Robert Evans and Nigel Wilding for insightful discussion and providing critical feedback on the ideas in this manuscript.  
PBW additionally thanks Andrew Masters for helpful discussions and drawing our attention to the literature on zero-separation theorems.
We also thank Rachel Hendrikse and Mark Wilson for providing the Monte-Carlo data from Ref.~\onlinecite{hendrikse_2025} for \Fig{fig:monomer-mu}.
\end{acknowledgments}

\appendix

\section{Other approaches to the solvent-mediated PMF}
\subsection{Test particle theory}\label{appendix:test-particle}
In the test particle framework, the excess part of the chemical potential is equated with the grand potential cost from inserting a trial particle:
\begin{equation}
\mu_i^\ex \equiv \Delta \Omega_i
\equiv \Omega\left[
\{ \rho_{j|i}(\rvec) \}
\right] - \Omega_\text{hom}\,,
\end{equation}
where $\Omega_{\text{hom}}$ is the homogeneous grand potential and $\rho_{j|i}(\rvec)$ is the density of species $j$ given there is an inserted trial particle of type $i$ which can be treated as an external potential.
For convenience the trial particle can be situated at the origin.

The grand potential change is decomposed as
\begin{equation}\label{eq:dOmega}
  \Delta \Omega_i = \Delta F_i^\id + \Delta F_i^\ex - {\textstyle\sum_j} {\textstyle\int} \drvec \, \psi_{j|i}(\rvec) \Delta \rho_j(\rvec)\,,
\end{equation}
where $F^\id$ is the ideal gas free energy and $\psi_j \equiv \mu_j - v_{ij}$ is the intrinsic chemical potential incorporating pairwise interactions $v_{ij}$ with the test particle.
The free energy due to interactions can be formally written as
\begin{equation*}
\begin{split}
\Delta F_i^\ex
=
- \kT \sum_{n,\kvec^n} \frac{1}{n!} \int \!\drvec_1\cdots \drvec_n \, &c_{k_1 \cdots k_n}(\rvec_1,\dots,\rvec_n) \\[-6pt]
&{}\times{\textstyle\prod_{j=1}^n} \Delta \rho_{k_j}(\rvec_j)\,,
\end{split}
\end{equation*}
where $\Delta \rho_j(\rvec) \equiv \rho_{j|i}(\rvec) - \rho_j$ is change in density from the reference homogeneous state $\rho_j$ in the absence of the test particle.
Here the direct correlation functions
\begin{equation}\label{eq:c-potential}
c_{k_1 \cdots k_n}(\rvec_1,\dots,\rvec_n) \equiv -\frac{\delta^n \beta F^\ex}{\delta \rho_{k_1}(\rvec_1) \cdots \delta \rho_{k_n}(\rvec_n)}
\end{equation}
are evaluated in the reference homogeneous state.

Equilibrium is found by minimising \Eq{eq:dOmega} through setting $\delta \Delta \Omega_i / \delta \rho_j = 0$.
After equating $\Delta \rho_{j|i} = \rho_j g_{ij} = \rho_j (1 + h_{ij})$, this results in the exact expression for $g_{ij}$ given in the main text \Eq{eq:gij-exact} with the bridge function defined as
\begin{equation}\label{eq:bridge}
\begin{split}
b_{ij}
=
\sum_{n\ge 3, \kvec^{n-1}} \frac{1}{(n-1)!}
\int\! &\drvec_1\cdots\drvec_{n-1} \,
c_{jk_1\cdots k_{n-1}} \\[-6pt]
&\qquad{}\times{\textstyle\prod_{\ell=1}^{n-1}} \Delta \rho_{k_\ell|i}
\end{split}
\end{equation}
accounting for many-body effects.
This formally exact theory is solved by neglecting the bridge function $b_{ij} \simeq 0$ and then two related theories emerge depending on how the indirect correlation $h_{ij} - c_{ij}$ is approximated.
The first is the hypernetted chain (HNC) where \Eq{eq:gij-exact} with $b_{ij} \simeq 0$ is used as a closure for the OZ relation in \Eq{eq:oz} which must then be solved self-consistently to find $h_{ij}$ and $c_{ij}$.
We will write the direct correlation that emerges from this route as
\begin{equation}\label{eq:c-hnc}
c_{ij}^\text{HNC}
=
h_{ij} - \ln{(1 + h_{ij})} - \beta v_{ij}\,.
\end{equation}
This is exactly \Eq{eq:gij-exact} with bridge function $b_{ij} = 0$.
The disadvantage of this route is that we have to determine $c_{ij}$ self-consistently, and thus possess no knowledge of the underlying free energy functional from its original definition in \Eq{eq:c-potential}.

Archer and Evans\cite{archer_2003} proposed an alternative approach where the random phase approximation (RPA) $c_{ij} \simeq c_{ij}^\text{RPA} = -\beta v_{ij}$ is applied inside the indirect correlation function.
This leads to
\begin{equation}\label{eq:quasi-oz}
h_{ij}
=
c_{ij}^\text{HNC}
+ {\textstyle\sum_k} \rho_k c_{ik}^\text{RPA} \convolve h_{kj} \,.
\end{equation}
This quasi-OZ relation bares a striking similarity to the true OZ relation in \Eq{eq:oz}.
This differs from the conventional way in which the RPA is applied, where both the $c_{ij}$ are replaced by $c_{ij}^\text{RPA}$ in \Eq{eq:oz}.
We will refer to this route as the test particle RPA (TP-RPA) to distinguish it from the conventional RPA route.
The advantage of approaches based on the RPA is that we \emph{can} write down the free energy functional \latin{a priori}, as
\begin{equation}\label{eq:fex-rpa}
 F^\ex[\rho(\rvec)] = {\textstyle\frac{1}{2}}\,{\textstyle\sum_{i < j}}\, {\textstyle\int} \drvec\, \drvec' 
 \rho_i(\rvec) \rho_j(\rvec') v_{ij}(|\rvec - \rvec'|)\,.
\end{equation}
It is generally a less accurate approximation than the HNC, which has a greater degree of self-consistency by construction, and becomes essentially exact for ultrasoft systems, such as the DPD potential in \Eq{eq:phi-dpd}, especially in the limit of high densities.

Under both the HNC and the TP-RPA approximations, the excess chemical potential is naturally obtained from $\Delta \Omega_i$ as
\begin{equation}\label{eq:mu-tp}
\beta \mu_i^\ex
\simeq
{\textstyle\sum_j}\,\rho_j
{\textstyle\int} \drvec \, [
{\textstyle\frac{1}{2}} h_{ij} ( h_{ij} - c_{ij}^\text{HNC} )
- c_{ij}^\text{HNC}]\,,
\end{equation}
\cf\ \Eq{eq:hncmui}.
For the RPA we can also write down the chemical potential via the direct compressibility route as
\begin{equation}\label{eq:mu-rpa}
\beta \mu_i^\ex
=
- c_i(\rvec)
\simeq
{\textstyle\sum_j}\, \rho_j {\textstyle\int} \drvec' \, \beta v_{ij}(|\rvec - \rvec'|)
\end{equation}
from the functional derivative of $F^\ex$ in \Eq{eq:fex-rpa}.
This is the conventional RPA approach which represents the genuine mean-field limit.
In their Eq.~(9), 
Archer and Evans\cite{archer_2003} considered the compressibility form but not the direct test particle expression as in \Eq{eq:mu-tp}.

The solvent-mediated potential $W_{\kvec}$ can be determined directly in the test-particle approach by inserting the grand potential change in \Eq{eq:dOmega} into the formally exact expression in \Eq{eq:w-exact}.
Alternatively, Archer and Evans\cite{archer_2003} recognised that the solvent mediated potential could be obtained in the infinite dilution limit from the one-body direct correlation of the inhomogeneous test particle system:
\begin{equation}\label{eq:wij-direct}
\beta W_{ij} = \beta \mu_j^\ex - c_{j|i}(r)
\end{equation}
where $c_{j|i}$ means $c_j$ evaluated under density profiles surrounding a test particle $i$ at the origin.
Under the RPA this gives
\begin{equation}\label{eq:wij-rpa}
\begin{split}
W_{ij}(r)
&=
\mu_j^\ex + {\textstyle\sum_k}\, {\textstyle\int} \drvec' \, \rho_{k|i}(\rvec') v_{jk}(|\rvec - \rvec'|)\\[3pt]
&\quad{}={\textstyle\sum_k}\, \rho_k  \,v_{jk} \convolve h_{ki} \,,
\end{split}
\end{equation}
with $k$ summing over solvent species.
This result is however not guaranteed to be symmetric under exchanging $i$ and $j$ if the particles are different.\cite{archer_2003}

\subsection{Reference Interaction Site Model (RISM)}\label{appendix:rism}
The reference interaction site model (RISM) consists of a generalised OZ equation,\cite{chandler_1982} which in reciprocal space is
\begin{equation}\label{eq:ozrism}
\Hq = \Omegaq\, \Cq\, \Omegaq + \Omegaq\, \Cq\, R\, \Hq\,.
\end{equation}
Here $\Hq$ and $\Cq$ are matrices characterising the correlations between atoms (or `sites') in \emph{different} molecules, and 
$R = \rho_1 \Id_1 \oplus \rho_2 \Id_2 \oplus \cdots \oplus \rho_n \Id_n$ where $\{\rho_i\}$ are 
the concentrations for each of the $n$ molecules and the $\{\Id_i\}$ are square identity matrices of dimension equal to the number of atoms in each molecule.
The RISM OZ equation can be formally derived from the atomistic OZ equation by defining $\Omegaq \equiv ( \Id - \Cq_{\text{in}} R )^{-1}$ to characterise the intramolecular correlations in terms of some intramolecular direct correlation function $\Cq_{\text{in}}$.

Specialising to the case of a dimer in a monomeric solvent, we label the solvent as `0', one end of the dimer as `1', and the other end of the dimer as `2'.
Then $\Omega$ has unit entries along the diagonal and the only non-zero off-diagonal entry is $\Omega_{12}$.
At infinite dilution, for the solvent one has $\hat{h}_{00} = \hat{c}_{00} (1 + \rho_0 \hat{h}_{00})$ so that with the HNC closure a one-component problem obtains as before.
For the solvent-site correlation functions one gets
\begin{subequations}\label{eq:rism}
\begin{align}
\hat{h}_{01} = \hat{h}_{10} &= (\hat{c}_{10} + \hat\Omega_{12} \hat{c}_{20}) (1 + \rho_0 \hat{h}_{00})\,, \\
\hat{h}_{02} = \hat{h}_{20} &= (\hat{c}_{20} + \hat\Omega_{12} \hat{c}_{10}) (1 + \rho_0 \hat{h}_{00})\,.
\end{align}
\end{subequations}
Unlike the monomer mixture case these are now coupled by the $\hat\Omega_{12}$ term and they have to be solved together.
For homodimers though, the system simplifies since we expect $c_{10} = c_{20}$ and $h_{10} = h_{20}$ by symmetry, giving
\begin{equation}\label{eq:rism-homodimer}
  \hat{h}_{10} = (1 + \hat\Omega_{12}) \hat{c}_{10} (1 + \rho_0 \hat{h}_{00})\,.
\end{equation}
This is now effectively equivalent to a one-component problem via a minor modification of the OZ relation in \Eq{eq:oz}, and can be solved as such with the HNC closure.

In the HNC-RISM approach the normal HNC expression for the chemical potential in \Eq{eq:mu-tp} is used for \emph{each} monomer.
The dimer chemical potential then follows as $\mu_1^\ex(\ell) + \mu_2^\ex(\ell)$.
In the infinite dilution limit only the correlations with the solvent matter, and the solvent-mediated PMF
is inferred from \cite{fraaije_2016}
\begin{equation}\label{eq:w11-rism}
W_{11}
=
\Delta \mu_1^\ex(\ell) + \Delta \mu_2^\ex(\ell)
=
2 \mu_1^\ex(\ell) - 2 \mu_1^\ex(\infty)\,.
\end{equation}
For rigid dimers of length $\ell$ the intramolecular correlations are taken as $\hat\Omega_{12} = \sin(q \ell) / (q \ell)$.

\section{Multicomponent solvents}\label{appendix:multi}
We indicate the appropriate generalisation of the main text theory to a multicomponent solvent.  First, 
write the OZ equations in matrix form as, \cf\ \Eq{eq:ozrism},
\begin{equation}
  \Hq = \Cq + \rho\Hq X\Cq\,,
\end{equation}
where $\Hq$ is the matrix of total correlation functions, $\Cq$ is likewise for direct correlation functions, $X$ is diagonal in the mole fractions, and $\rho$ is the total density.
For the case of solutes in a mixed solvent, we partition the matrices into solvent and solute sectors, denoted by `0' and `1' respectively, and take the limit of infinite dilution for the solutes.
This yields
\begin{subequations}
  \begin{align}
      \Hq_{00} = \Cq_{00} + \rho_0\Hq_{00} X_0\Cq_{00}\,,\\
      \Hq_{01} = (I + \rho_0\Hq_{00} X_0)\Cq_{01}\,,\\
      \Hq_{11} = \Cq_{11} + \rho_0\Hq_{10} X_0\Cq_{01}\,,
  \end{align}
\end{subequations}
(omitting the lower left quadrant, which duplicates the upper right). The first of these are the standard OZ equations in the solvent sector.  To make sense of the remainder, let us introduce the partial structure factor matrix in the solvent\cite{hansen_2006}
\begin{equation}
  S(q)=X_0+\rho_0 X_0 \Hq_{00} X_0\,,
\end{equation}
or in component form
\begin{equation}
  S_{\mu\nu}(q)=x_\mu\delta_{\mu\nu}+\rho_0 x_\mu x_\nu {\tilde h}_{\mu\nu}(q)\,.
\end{equation}
(other normalisations are possible, but we stick with this one here). It follows from the definition that $S$ is symmetric.  Hansen and McDonald~\cite{hansen_2006} show that $N S_{\mu\nu}(k) = \langle {\tilde\rho}_\mu(\kvec) {\tilde\rho}_\nu(-\kvec) \rangle$ where $N$ is the number of (solvent) particles.  Then, if $y_\mu$ is any vector, 
\begin{equation}
  N{\textstyle\sum_{\mu\nu}} y_\mu y_\nu S_{\mu\nu}
  = \langle|{\textstyle\sum_{\mu}}y_\mu{\tilde\rho}_\mu(\kvec)|^2\rangle > 0\,.
\end{equation}
This demonstrates that $S$ is also positive definite.  Hence the eigenvalues are all positive, and the matrices $S^{-1}$ and $S^{-1/2}$ exist and are well defined.

In these terms the off-diagonal block of the OZ relations becomes $\Hq_{01}=X_0^{-1}S\Cq_{01}$.  This can be closed with the HNC approximation to determine the mixed solvent-solute correlation functions $\Hq_{01}$.  Importantly, it also shows $\Cq_{01}=S^{-1}X_0\Hq_{01}$.  Inserting into the solute sector OZ equations shows that
\begin{equation}
  -\Hq_{11}+\Cq_{11}=-\,\rho_0\,\Hq_{10}X_0S^{-1}X_0\Hq_{01}\,.
\end{equation}
Assuming the HNC closure, we identify $W_{11}=-H_{11}+C_{11}$ as the matrix of solvent-mediated PMFs.  We now cast this as
\begin{equation}
  \tilde W_{11} = -\,p_0\, {\tilde\Psi}_{10} {\tilde\Psi}_{01}\,,
\end{equation}
where $p_0$ is the solvent pressure, and
\begin{equation}
{\tilde\Psi}_{10}={\tilde\Psi}^{\,\transpose}_{01} = \sqrt{\rho_0/\beta p_0}\times S^{-1/2} X_0 \Hq_{01}\,.
\end{equation}
This generalises the construction in \Eqs{eq:psi-def}--\eqref{eq:wij-general} in the main text single-component analysis.  In component terms, for every solute bead type $i$, $j$ etc, and solvent bead type $\mu$, $\nu$ etc, construct
\begin{equation}
  {\tilde\psi}_{\mu i}=\sqrt{\rho_0/\beta p_0}\times{\textstyle\sum_\nu}\>
  [S^{-1/2}]_{\mu\nu}\,x_\nu \,{\tilde h}_{\nu i}\,.
\end{equation}
Then the solvent mediated PMF between solute bead types $i$ and $j$ is a sum of convolutions,
\begin{equation}
  W_{ij}(r)=-\,p_0\,{\textstyle\sum_\mu}\>\psi_{\mu i} \convolve \psi_{\mu j}\,.
\end{equation}
Note that the sole structural input from the solvent here is the matrix of partial structure factors.  In the present scheme it is envisaged that these come from the HNC closure of the solvent OZ equations, but in principle they could be taken from another approximation, or even measured in simulation.

\bibliography{selected}

\end{document}